\documentclass[12pt]{article}
\usepackage{amsmath}
\usepackage{times}
\usepackage{graphicx}
\usepackage{color}
\usepackage{multirow}
\usepackage[margin=1cm,labelfont=bf]{caption}

\usepackage{rotating}
\usepackage{bbm}
\usepackage{latexsym}

\usepackage{hyperref}
\usepackage[outdir=./]{epstopdf}
\usepackage[numbers]{natbib}

\begin{document}
{\scriptsize \emph{Author's version.} Published in Neural Computation 2017 29:9, 2491-2510 \href{http://dx.doi.org/10.1162/neco_a_00987}{\nolinkurl{doi.org/10.1162/neco_a_00987}}}\\

\begin{center}
{\LARGE Cortical Spike Synchrony as a\\ Measure of Input Familiarity}
 \end{center}
\ \\
{\bf Clemens Kornd\"orfer} \hfill ckorndoe@uni-osnabrueck.de\\
{\bf Ekkehard Ullner$^{\displaystyle 2}$} \hfill e.ullner@abdn.ac.uk\\
{\bf Jordi Garc\'{i}a-Ojalvo$^{\displaystyle 3}$} \hfill jordi.g.ojalvo@upf.edu\\
{\bf Gordon Pipa$^{\displaystyle 1}$} \hfill gpipa@uni-osnabrueck.de\\
\ \\
{$^{\displaystyle 1}$Institute of Cognitive Science, University of Osnabr\"uck, Osnabr\"uck, Germany.}\\
\ \\
{$^{\displaystyle 2}$Department of Physics, Institute for Complex Systems and Mathematical Biology and
Institute of Medical Sciences, University of Aberdeen, Aberdeen, United Kingdom.}\\
\ \\
{$^{\displaystyle 3}$Department of Experimental and Health Sciences, Universitat Pompeu Fabra, Barcelona Biomedical
Research Park, Barcelona, Spain.}\\

\
\begin{center} {\bf Abstract} \end{center}
Spike synchrony, which occurs in various cortical areas in response to specific perception, action and memory tasks, has sparked a long-standing debate on the nature of temporal organization in cortex. One prominent view is that this type of synchrony facilitates the binding or grouping of separate stimulus components. We argue instead for a more general function: A measure of the prior probability of incoming stimuli, implemented by long-range, horizontal, intra-cortical connections. We show that networks of this kind -- pulse-coupled excitatory spiking networks in a noisy environment -- can provide a sufficient substrate for stimulus-dependent spike synchrony. This allows a quick (few spikes) estimate of the match between inputs and the input history as encoded in the network structure. Given the ubiquity of small, strongly excitatory subnetworks in cortex, we thus propose that many experimental observations of spike synchrony can be viewed as signs of input patterns that resemble long-term experience, i.e. patterns of high prior probability.

%%%%%%%%%%%

\section*{Introduction}

Depending on the behavioural context, specific cortical neurons fire in synchrony. In the sensory cortex, this depends on qualities of the sensory input: Sounds evoke simultaneous activity in auditory cortex cells with matching receptive fields \citep{brosch2002,atencio2013}, distant cells in somatosensory cortex synchronize when particular skin regions are stimulated \citep{reed08}, and synchrony in primary visual cortex (V1) varies with geometrical stimulus features such as spatial continuity \citep{gray89, living96} or similarity of orientation \citep{kohn05}. This can be observed as soon as 30 ms after a stimulus change \citep{maldonado}, i.e. within only a few spikes. Beyond sensory cortex, spike synchrony in primary motor cortex varies with either the performed \citep{jackson2003} or the intended action \citep{riehle97,pipa2006} and it has has been shown that distant prefrontal cortex cells synchronize selectively during memory tasks \citep{pipa2011}.\\
Interpretations of such results have been conflicting. They have either been seen as an (inconsequential) epiphenomenon of cortical connectivity \citep{shadlen99}, or as evidence for a synchrony-based mechanism to quickly assemble and group different sources of currently relevant information \citep{roelfsema96,gray99}. Regardless of their function, the neural basis of these effects is mostly sought in long-range, horizontal, intra-cortical connections \citep{stettler02,ligilbert2002}.\\
Horizontal connections are known to adapt to experience both during development \citep{galuske96,schmidt99} and in adult learning \citep{rioultpedotti}. In the case of V1, their structure has been found to reflect the aggregate statistics of natural visual scenes \citep{onat13}. Perhaps as a result of such adaptations, the fine-scale topology of these networks is complex and functionally heterogenous \citep{martin14,rothschild2010}, i.e. connections between cells with differing response properties are common. However, one of the few, broad regularities in these networks appears to be that cells that respond in similar contexts tend to be well connected. This holds across many cortical areas. For example, direct horizontal connections are found between distant auditory cortex cells with similar response selectivity \citep{read2001} and between primary motor cortex cells representing related muscle groups \citep{weiss94}. V1 cells with nearby receptive fields are preferentially connected, even more so where they select for similar visual orientations \citep{tso86,stettler02}. V1 connectivity may further favour cells whose receptive fields fall in line with the axis given by their orientation tuning (\citet{bosking97}, but see \citet{martin14}). In somatosensory cortex, horizontal connections occur between distant cells that respond to sensations at opposing finger tips \citep{negyessy2013}. This illustrates that horizontal connectivity is not simply determined by receptive field similarity, but more generally seems to favour cells that are activated jointly in common sensorymotor contexts (such as handling an object between two fingers or seeing a continuous line). Conversely, thus, these are contexts which activate well-connected groups of cells.\\
To provide a concrete illustration, we briefly turn to visual cortex. V1 cells are known to selectively respond to stimuli with retinal coordinates matching their cortical position (retinotopy) and with a particular angle (orientation tuning). Consider a visual pattern formed by various short line segments. In case these elements are scattered across the visual field, retinotopy implies that a spatially scattered set of cells is activated. Few of these cells will have direct connections, since horizontal connections preferably connect cells with nearby receptive fields. More generally thus, the shortest path between any two responding cells will likely be longer, on average, for such a scattered stimulus than for a more compact stimulus.
A similar illustration can be made with respect to orientation: Consider a chain of line segments. In case their orientations are aligned, a spatially neighbouring set of cells with similar orientation tuning responds. The cited patterns in V1 connectivity suggest denser connectivity between these cells than between a -- more heterogenously tuned -- set of cells that would respond to a chain of heterogenously oriented segments. Again, depending on the involved distances, this may translate to a difference in the length of paths between the activated cells. A much more complete, idealized, geometrical model of V1 horizontal connection patterns in relation to visual grouping has been provided by \citet{benshahar}.\\
In summary, it appears that cortex in certain situations receives activation patterns that ``match'' the existing lateral network structure particularly well, in the sense that they activate closely-connected groups of cells. Further, the network structure appears to reflect experience in the sense that commonly co-activated units are better connected. Thus, to show that synchrony reflects the similarity of an incoming spatial pattern to the patterns commonly seen in the past, one primarily needs to show that synchrony reflects the degree to which the current network connects the cells activated by that pattern. The question therefore is why, in this case, there is an increase in spike synchrony.\\
It seems unlikely at first sight that the discussed horizontal network by themselves explain the synchrony effect at hand: These connections establish an excitatory \citep{tso86,sincich01}, pulsed (chemical) coupling of spiking cells \citep{mcguire91}. In theoretical models of such networks, synchrony, if possible at all, has been found to be unstable in the sense that the phases of different cells will spontaneously align and disperse in reaction even to small disturbances \citep{ernst98}, depending on delay durations, connection strengths and the network topology \citep{perez11}.\\
Consequently, at the heart of all existing spiking network models of cortical synchrony are oscillators formed by the interaction of excitatory and inhibitory cells, in which the latter counterbalance the destabilizing effects of excitation (see e.g. \citet{wilson91,boergers2005}). Here, we are going to argue that the reported findings of cortical synchrony by themselves do not require or imply the presence of excitatory-inhibitory oscillators.\\
We mentioned that an important aspect of cortical synchrony is its dependence on spatial features of incoming stimulus patterns. The model by \citet{wilson91} is notable for replicating one such effect \citep{gray89} in an excitatory-inhibitory spike based model. Richer stimulus-dependent synchrony effects are found in a group of models that abstract away from the (spiking) dynamics of biological networks, aiming for an interpretation of cortical synchrony as a binding mechanism. Specifically, these models demonstrate the emergence of synchronous cell assemblies that reflect large-scale, spatial relations within a stimulus, which for example leads to an elegant, ``self-organized'' solution for visual grouping or image segmentation problems \citep{li98,yenfinkel98,schillen94,wang95local,wang95global,finger14}. These synchronization effects seem to rely on properties of continuously coupled rate- or phase models, as they have not yet been demonstrated in pulse-coupled networks of spiking cells. A further, more exotic approach to achieve synchronous cell assemblies is to reconfigure the network structure for each new stimulus pattern \citep{yenfinkel98, wang95local, yuslotine}, though this is biologically motivated in only one case \citep{wang95local} to a certain degree. Finally, it is interesting to note that \citet{vanrullen} criticized the whole category of models cited here as implausibly slow for the purpose of stimulus grouping and proposed a fast, feed-forward, spike-based grouping model, without regard to synchronisation.\\
Here, in an attempt to find a minimal model that has the required computational properties, we provide conditions under which fixed, purely excitatory, chemically coupled spiking networks alone exhibit the required, fast stimulus-dependent synchrony response. One relevant property of such networks is that they can produce temporally ordered spike responses in the presence of noise. Perhaps counterintuitively, noise appears as a beneficial factor in various biological contexts, such as improved neural signal transmission or increased spike time reliability \citep{ermentrout08,abbot09}. In particular, excitable systems -- systems such as spiking neurons, whose response to a sufficient input is stereotypical and followed by a refractory period -- are known to engage in periodic oscillations when they receive noise input of some optimal, non-zero amplitude \citep{lindner04}. This effect (coherence resonance or autonomous stochastic resonance) has been shown in the FitzHugh-Nagumo \citep{pikovsky}, Hodgkin-Huxley \citep{lee98} and Morris-Lecar \citep{balenzuela05} neuron models and it extends to arrays of multiple, sufficiently coupled spiking neurons, i.e. to excitable media \citep{neiman99}. Such networks, when driven by random input, can display regular traveling waves, among other spatially organized activation patterns. We build on these findings and show that this type of emergent, spatial organization of synchronous activity in excitatory networks allows a fast synchrony-based measure of the ``match'' between incoming spatial activation patterns and the current network topology.\\
This article is structured as follows. We first show that random spike input is transformed to synchronous responses in various excitable, pulse-coupled networks, in particular when stimulating well-connected subpopulations. This basic effect is then used to demonstrate synchrony as a measure of similarity between incoming stimuli and the long-term stimulus history as reflected in current network structures (``familiarity''). Additionally, we show that the effect is compatible with Gray et al.'s observation of stimulus dependent synchrony \citep{gray89} and provide some conditions for its occurence, and close with a comparison to known cortical dynamics.

\begin{figure}[h]
    \centering
    \includegraphics[width=0.7\textwidth]{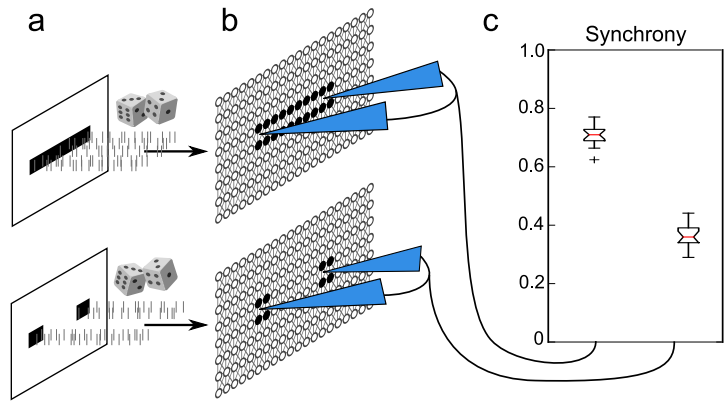}
    \caption{\textbf{Basic setup.} (a) A stimulus pattern is defined and uncorrelated random spike trains are sent to the subset of cells specified by the pattern (black cells, b). (b) Activity in the excitatory network is measured in various locations -- here, two fixed measurement sites shown in blue. (c) The degree of zero time lag synchrony of the measured locations varies with the stimulus pattern.}
\label{fig:overview}
\end{figure}

\section*{Results}
\begin{figure}[h!]
    \includegraphics[width=1\textwidth]{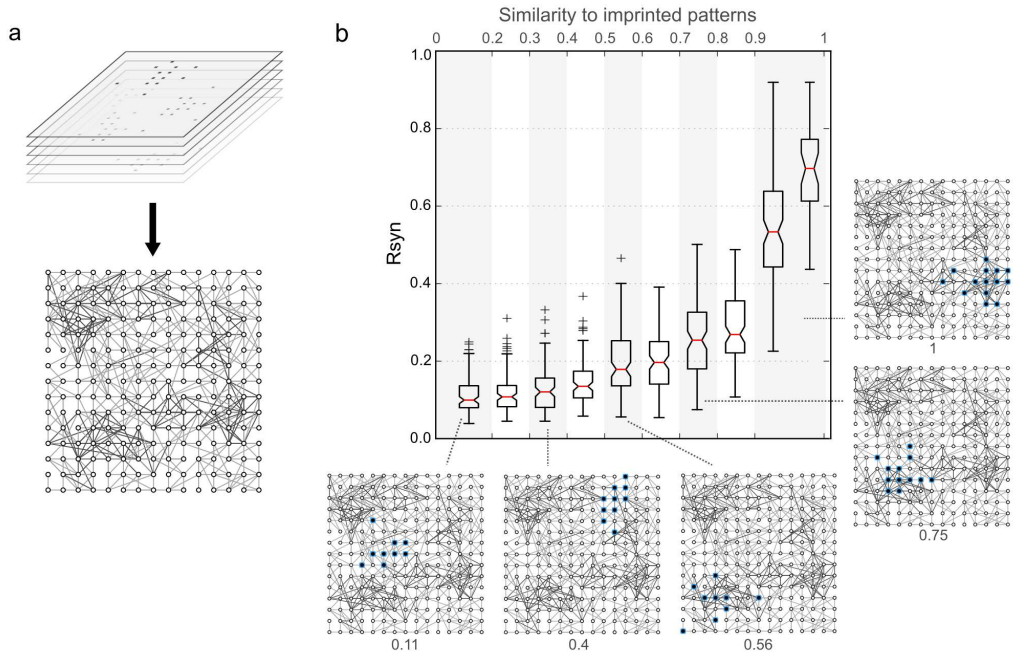}
    \caption{\textbf{Synchrony as a familiarity measure.} (a) We sample spatial networks with a distance-dependent connection probability. For each, a set of random stimulus patterns is fixed, and stronger connectivity is allowed to occur between cells co-activated within such patterns. (b) For each network, many new stimulus patterns are sampled and binned by their similarity (degree of overlap) to the set of patterns ``imprinted'' in the previous step. The box plot shows the degree of zero time lag spike synchrony for 100 samples from each similarity bin, taken across networks and patterns. The final bar shows the subset of the penultimate bin where two or more connections per activated cell have occured. Insets: Samples of input patterns of different degree of similarity to the imprinted pattern set of one exemplary network.}
\label{fig:familiarity}
\end{figure}

\begin{figure}[h!]
    \centering
    \includegraphics[width=0.7\textwidth]{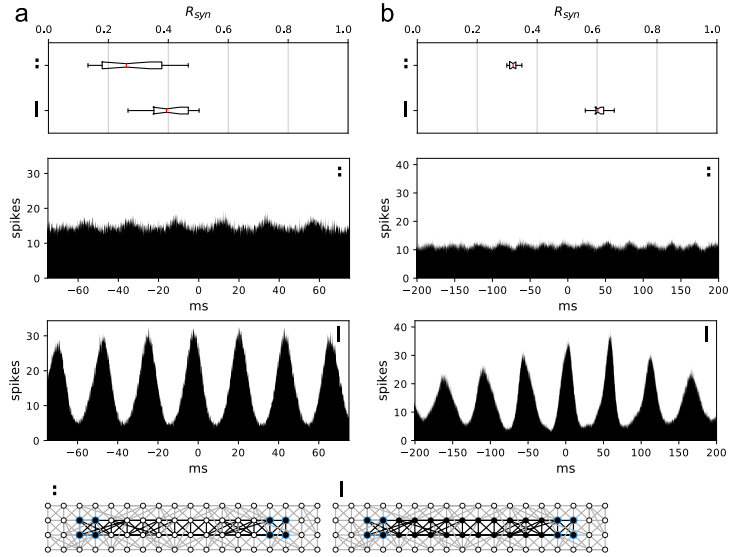}
    \caption{\textbf{Spike synchrony reflects global stimulus properties.} Two stimulus patterns were compared on a fixed network set in the coherent spiking (a) and the coherent chattering regime (b). Top panels show the degree of zero-time lag synchrony of two fixed measurement populations. Middle panels show cross-correlograms between the two measured sites for the two stimuli. Bottom insets show the two stimulus patterns (black markers) and fixed measurement sites (blue markers).}
\label{fig:bridge}
\end{figure}

We study networks of single-compartment Izhikevich spiking neurons \citep{Izhikevich_neuron} with excitatory chemical (pulsed) coupling. A subpopulation receives uncorrelated, stationary, random (Poisson) input spike trains on excitatory and, optionally, inhibitory synapses. As the network responds, the degree of zero time lag spike synchrony of a particular group of cells is measured. This basic setup is shown in Figure \ref{fig:overview}.\\
\begin{figure}[t]
    \includegraphics[width=1\textwidth]{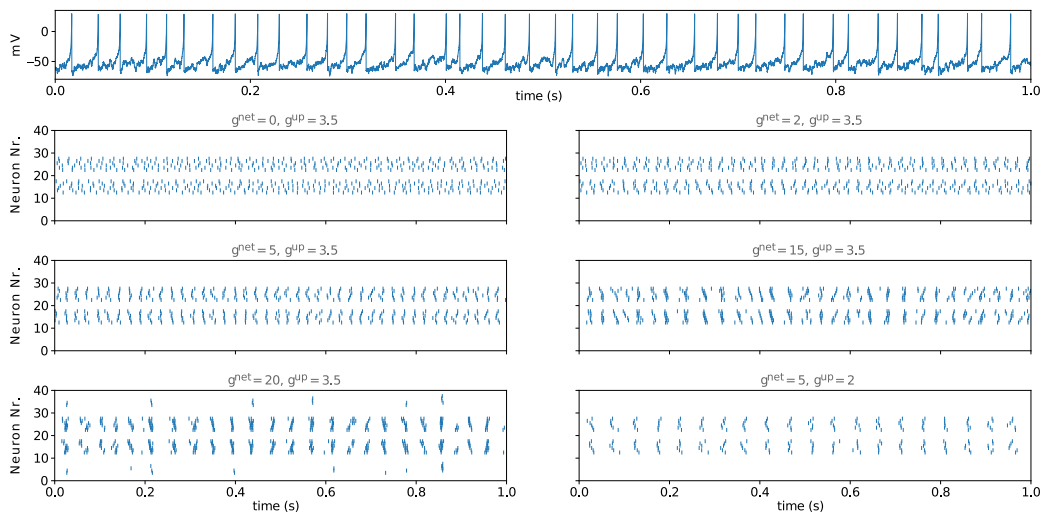}
    \caption{\textbf{Model behaviour.} One second of activity. Top panel: Voltage trace of a single unit. Other panels: Spike activity in a grid network receiving input in a central subpopulation, for increasing lateral synapse conductances $g^{net}$ from left to right and top to bottom. (The bottom right panel shows a lower upstream input conductance $g^{up}$.) Networks shown unrolled row-by-row along the vertical axis.}
\label{fig:behaviour}
\end{figure}
The model neuron is set to a regime of integrating, Class 1 behaviour, firing at a much lower rate than the rate of external input pulses (a feature shared by many cortical neurons \citep{shadlen98}). Depending on connectivity, networks of such neurons can generate synchronous responses. Figure \ref{fig:behaviour} illustrates this by showing the spike activity in a 4 x 10 lattice network with eight-nearest-neighbour connectivity, receiving random input in a central, fixed subpopulation of 12 units. The different panels show the network's response given increasing lateral synapse conductances $g^{net}$. Coherent spiking emerges already at lower conductances, followed by a regime of coherent chattering, followed at higher conductances by a regime of spreading activity beyond the input-receiving population. The first two regimes are considered in the following.\\
Animations of network activity over time (see online\footnote{\href{https://youtu.be/DiBIYmj_9DU}{\nolinkurl{youtu.be/DiBIYmj_9DU}}, \href{https://youtu.be/Tx5CbLyJutQ}{\nolinkurl{youtu.be/Tx5CbLyJutQ}} and \href{https://youtu.be/EEi5Fg2YOtE}{\nolinkurl{youtu.be/EEi5Fg2YOtE}}}) show that a single oscillation cycle is characterized by one or more waves of activation spreading out from one or more (roughly simultaneously appearing) origin points, passing through the input-receiving region until extinction occurs either by collision with another wave or by reaching the stimulus boundary. Fewer, broader waves are observed in networks with strong lateral connectivity. In short, the input-receiving region appears to behave as an excitable medium during the presence of a stimulus.\\
\begin{figure}[t]
    \includegraphics[width=1\textwidth]{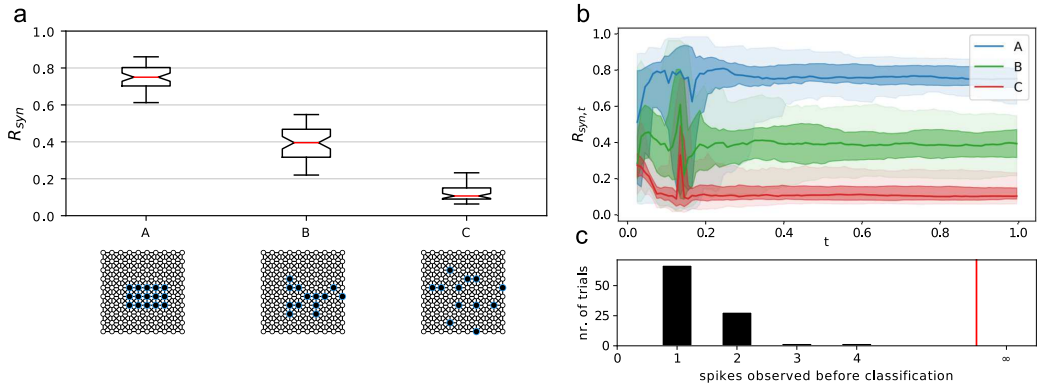}
    \caption{\textbf{Synchrony reflects the match between input patterns and network structure.} (a) Subpopulations with different average path lengths on the given network are driven by random spike inputs (black dots) as the synchrony of the activated cells is measured. (b) Synchrony measured in growing windows. Middle lines denote the median, coloured bands the two middle and two outer quartiles, of 50 trials where synchrony was measured in intervals from $t=0$ to the corresponding time on the lower axis. (c) Speed of synchrony divergence between stimulus A and C. The histogram shows in how many trials these conditions could be successfully classified by synchrony after observing a certain average number of spikes. Here, the conditions could be discerned by the synchrony of the first spike wave in most trials.}
\label{fig:dispersion}
\end{figure}
In general, different stimulus patterns -- different choices of input-receiving cells -- lead to different degrees of average zero time lag synchrony ($R_{syn}$) among these cells, depending on the connections that exist between them. This is best visualized on a lattice network. Figure \ref{fig:dispersion}a shows such a network set in the chattering regime ($g^{net}=15$). For a stimulus that activates cells connected by short paths, these cells fire more synchronously than the more dispersed cells activated by a scattered version of the stimulus.\\
Synchrony differences between different stimuli are discernible after a few spikes. Figure \ref{fig:dispersion}b shows synchrony measurements taken over increasingly long time windows (each starting from stimulus onset at $t=0$).\\
To quantify the information content of such measurements, we compute the length of time for which the network needs to be observed until a naive Bayes classifier will correctly infer which stimulus was presented, based on the measured synchrony values. (See Methods for details). We report the number of spikes that need to occur, on average over the measured set of cells, until such a classification succeeds (histogram panels in Figures \ref{fig:dispersion}, \ref{fig:bridge}). For example, in case of the three stimuli in Fig \ref{fig:dispersion}, it is in almost all trials sufficient to observe the degree of synchrony of one or two spike waves. While simple grid networks as shown have the advantage of allowing an easy visualization of topological distances, we note that the same effect also occurs on other network topologies, such as networks where connection probability diminishes with spatial distance and `small world' networks with power-law degree distribution (see supplemental material).\\
We showed excitatory spiking networks in which well-connected subpopulations fire synchronously when they receive (noisy) external drive. Above, we argued that behaviour like this is pivotal in explaining the interplay of cortical response selectivities, learned excitatory horizontal network structures and experimental observations of synchrony. Further, a connection to stimulus familiarity is implied: Since cortical horizontal connections concentrate between cells that have often been co-activated by previous stimuli, a stimulus that activates well-connected cells is likely similar to (parts of) these past stimuli, i.e. familiar. To make this connection somewhat more explicit, in Figure \ref{fig:familiarity} we turn to networks with more heterogenous structure, shaped by stimulus patterns.\\
We sample networks with random, local connectivity. Specifically, the probability that two different cells $u,v$ on a $15\times15$ integer lattice are connected is set to fall with their euclidian distance, $P_{connect}(u,v) = max(0,d(u,v)^{-1} - c)$. The cutoff $c$ varies between cell pairs: For each network, a set of random stimulus patterns plays the part of long term stimulus history, i.e. these patterns are taken to have high prior probability (see also Methods). In sampling the network, $c$ is then relaxed (from $c=0.3$ to $c=0.15$) between any cells that co-occur in such a pattern (allowing slightly longer links between such cells). These connections are also stronger ($g^{net} = 15$ vs. $1$ elsewhere). As a result, several of the cell pairs co-activated in such patterns end up being directly and strongly connected. In other words, a number of strong excitatory subnetworks have been embedded in the larger network structure (concentrated around certain stimulus configurations), leading to a network that, under suitable external input, can \emph{partially} behave as an excitable medium. The network structure is then kept fixed, since we are only concerned with network dynamics during short stimulus presentations. A model of the origin of such clustered network structures in terms of an interaction of plasticity rules and axonal delays has been proposed by \citet{izhikevich}.\\
The stimulus patterns are random and local, in the sense that the probability that a cell is counted as active in a certain pattern falls with the distance (like above) from a -- randomly chosen -- center point. After generating the network, additional patterns are sampled to serve as new, incoming stimuli. We define the ``familiarity'' of such a new pattern as the fraction of its cells shared with the initial set of patterns that has shaped the network (its similarity to these patterns). This can also be expressed as a prior probability, see Methods. We define ten intervals (bins) for this familiarity value and, by rejection sampling, generate a number of stimulus patterns in each bin.
For generality, the whole sampling procedure of networks and patterns is repeated until 100 examples (different networks with different input patterns) have been drawn in each bin. There are thus three main sources of variance in the synchrony measurements displayed in Figure \ref{fig:familiarity}: First, the network sampling procedure is stochastic, meaning that nominally ``familiar'' patterns do not necessarily materialize in the form of a well connected network. To emphasize this factor, the last bar in Figure \ref{fig:familiarity} shows the subset of trials in which at least two connections have been sampled per stimulated cell, i.e. the subset of trials in which the network clearly reflects the pattern. Second, variance is introduced by the random initial patterns that define the network structure, which may for example overlap. Third, there is some degree of trial-by-trial variability of any single pattern on a single network as seen in Figure \ref{fig:dispersion}. \\
Despite these factors, an increase in synchrony is apparent as we increase the similarity of incoming stimulus patterns to the patterns embodied in the network structure.\\
This result can be explained in terms of the effect shown before: When an incoming pattern happens to hit an excitable (strongly connected) subnetwork, dynamics as in Figure \ref{fig:dispersion} play out. There, we have shown that the better a group of activated cells is connected by such strong connections, the higher its synchrony. Here, as in cortex, these connections are clustered around certain familiar stimulus patterns, hence stimuli that resemble these patterns produce stronger synchrony.
Note that this is not simply a matter of activating cells in the vicinity of strong connections: In Figure \ref{fig:familiarity}, the pattern with similarity index 0.56 (bottom row) is located in a relatively strongly connected region, but evidently misses part of the relevant subnetwork.\\
\newpage
In the cited experimental reports of cortical synchrony, only a few \mbox{(multi-) electrode} recording sites are usually set, which stay fixed throughout the various stimulus presentations. Mimicking this scenario, we find that two fixed, small groups of cells will fire with increased synchrony if a stimulus activates cells on the direct path between those groups (Figure \ref{fig:bridge}). This network was generated by the same procedure used in Figure \ref{fig:familiarity}, with increased connectivity concentrated in a $2\times14$ horizontal line (as would be expected e.g. in a group of similarly orientation-tuned V1 cells often exposed to continuous contours). Lateral conductances are $g^{net} = 5$ (a) and $g^{net} = 15$ (b).\\

\begin{figure}[t]
    \includegraphics[width=1\textwidth]{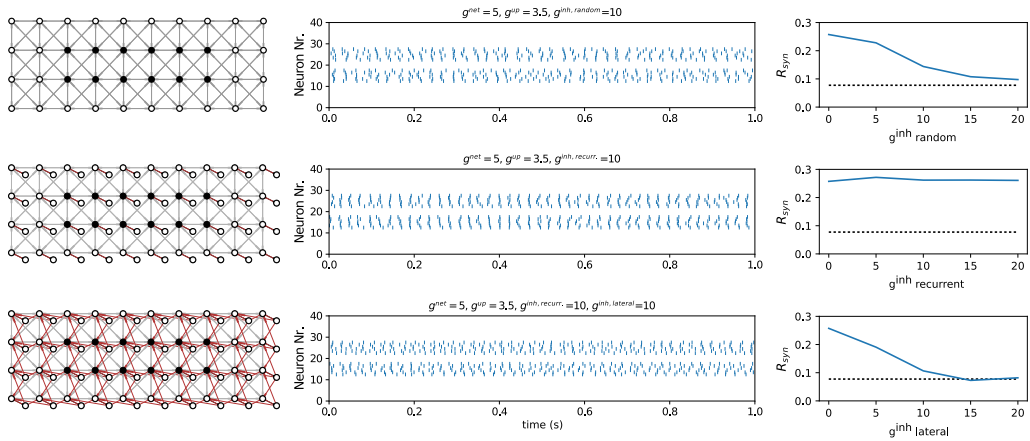}
    \caption{\textbf{Influence of various types of inhibition.} Uncorrelated inhibitory input (top), pairwise recurrent inhibition (middle), recurrent and lateral inhibition (bottom), for inhibitiory conductances ranging from zero to four times the excitatory lateral conductance. Left to right: Network visualization with inhibitory connections in red, connection targets marked by thicker line ends. Exemplary spike activity. Synchrony vs. inhibitory conductance -- dotted line indicates the degree of synchrony in the unconnected network.}
\label{fig:inh}
\end{figure}

The displayed effects are robust to interfering inhibitory input of various temporal structures. Figure \ref{fig:inh} shows the same network as Fig. \ref{fig:behaviour} under three different settings: Under uncorrelated inhibition, the stimulus-driven cells receive random inhibitory input spikes in addition to the excitatory drive. As the conductance of inhibitory input synapses increases beyond that of the lateral excitatory connections, synchrony begins to gradually decline. Under pure feedback inhibition (in which each excitatory cell projects to an additional inhibitory interneuron which projects directly back to it), even strong inhibitory conductances have little effect on synchrony. This changes with the addition of lateral inhibition (in which the inhibitory interneuron projects not only back to its driving excitatory cell, but also to the neighbours of that cell). Here again, we see a gradual decline in synchrony as inhibitory conductances increase beyond the strength of the lateral excitatory network.
In sum, we find that the presence of moderately strong inhibition of various temporal structures is not sufficient to destabilize the discussed synchronous dynamics, but offers a mechanism to dampen them gradually.
\clearpage

\section*{Discussion}
We showed examples of noise-driven, pulse-coupled spiking networks that display synchronous responses to random pulse inputs. By directing this input to cells that are closely and strongly connected (or located on a direct strong path between measured network locations), synchrony is increased. In accordance with experiments \citep{maldonado}, synchrony diverges quickly between different stimuli, i.e. within few spikes. \\
Above, we motivated our choice of network model by the properties of cortical long-range horizontal connections: These are distance-limited, excitatory, chemical (pulsed) couplings between spiking cells, with a topology shaped by long-term experience in the sense that commonly co-activated cells are connected more closely, and much more strongly \citep{cossell}.
But independent, excitatory subnetworks of functionally related cells also exist at smaller scales \citep{yoshimura} -- in fact, since plasticity in cortical networks seems to encourage cluster formation \citep{izhikevich}, such subnetworks may be ubiquitous. For instance, layer 5 cortical microcircuits appear to form a network of small, strongly excitatory subnetworks \citep{song2005}. The presented dynamics may therefore be found in structures other than the horizontal connections discussed, the more so as these dynamics occur across a broad range of synapse strengths, noise rates, and diversity of network topologies.\\
The discussed effect appears to be linked to the presence of traveling spike waves within the input-receiving population. This does not imply that the model predicts large wave fronts traveling across cortex: Given that strong synaptic connections (on which such waves depend) tend to be highly clustered, spreading waves may play out entirely within relatively confined subnetworks, undetectable at the mesoscopic scale. In other words, the proposed excitable dynamics are compatible with situations where no large-scale traveling fronts are observed. Such fronts do however occur in certain situations \citep{sato12,takahashi2015}: In visual cortex, they are mostly found during presentations of small, isolated stimuli \citep{sato12,neuhaus} and their size and number appears to peak directly following a stimulus change, after which they ease off progressively \citep{wessel}.\\
Inhibitory interneurons play no part in creating the presented synchrony effects; conversely, the effect is robust to moderate levels of interfering inhibitory input, which gradually attenuates it. Reports of cortical stimulus-dependent spike synchrony thus do not, by themselves, imply the presence of excitatory-inhibitory oscillator pairs.\\
We have isolated excitable dynamics as a common mechanism to explain a number of reports of cortical spike synchrony, including some of the experiments that sparked the binding-by-synchrony debate.
Functionally, this leads to a quick, locally computable, spike timing-encoded measure of stimulus familiarity (or more generally, beyond sensory areas, of the familiarity of a given spatial activation pattern): As cited above, horizontal connectivity patterns across cortex reflect long-term experience in the sense that commonly co-activated units are particularly well connected. In an excitable dynamic regime as presented here, higher spike synchrony is thus expected for input patterns that resemble experience. Hence, for example, synchrony may increase during perceptions of coherent visual stimuli such as connected lines simply because these are often experienced structures in our visual environment, which the network structure reflects. In such early sensory contexts, familiarity may thus express the degree to which a stimulus is structure or noise: Most behaviourally relevant objects have a highly structured appearance (consisting for example of Gestalt-like unbroken lines), while other, more noisy signals (with fewer spatial correlations) are typically found in less urgently important phenomena, such as background textures. In other words, often-experienced, behaviourally relevant stimuli are characterized by high mutual predictability of their different constituent parts and synchrony appears to signal this \citep{VinckBosman}. The result is a receptive field-like effect on the network level, in which a given group of cells is “tuned” to certain, well-correlated spatial arrangements of incoming activity and responds with a particular activation signature to these patterns but not to others, constituting a feature extraction step that operates on the level of groups of activated cells.\\
In more abstract terms, we have proposed that various cortical networks have access to a synchrony-encoded estimate of the prior probability of observing the current input pattern. A first estimate is available directly after the onset of the pattern (since synchrony in the first few spike waves is often already informative), after which precision continually improves. Hence, such a signal could be used early after input onset in a feed-forward fashion, for example to guide attention towards stimuli composed of plausible parts. More generally, estimates of prior probabilities are a prerequisite in Bayesian accounts of perception and learning, but it is unclear how such probabilities are represented neurally \citep{fiser}. We suggest that a spike-based encoding with the presented mechanism allows to rapidly generate and transmit such signals.

\section*{Methods}
\subsection*{Network model}

The two-dimensional Izhikevich spiking neuron model is a simplification of the Hodgkin-Huxley model of membrane conductances, but has a comparable dynamic repertoire \citep{Izhikevich_neuron}. Each neuron $i$ is given by its membrane potential $v_i$ and recovery variable $u_i$:
\begin{eqnarray}
\dot{v_i} &=& 0.04 v_i^2 + 5v_i + 140 - u_i - I_i^{net} - I_i^{up}\label{eq:izhi1} \\
\dot{u_i} &=& a(bv_i - u_i) \label{eq:izhi2}
\end{eqnarray}
Here we consider $a=0.01$, $b=-0.1$. Spikes are discrete events, triggering a reset of the model: Upon crossing the spike detection threshold of $v_i \geq 30mV$, $u_i$ is increased by 12 and $v_i$ is set to $-65mV$. The neuron receives input currents both from other cells in the network ($I_i^{net}$) and from external, ``upstream'' sources ($I_i^{up}$). These currents are, each, the sum of a number of individual synaptic currents that evolve according to the nonlinear, chemical synapse model proposed by \citet{destexhe94} (see also \citet{balenzuela05}). Intuitively, these synapses cause quickly increasing currents in response to incoming spikes, diminishing somewhat more gradually back to zero if no additional spikes arrive. A more specific description follows, beginning with the excitatory input the neuron receives from its neighbours in the network:
\begin{eqnarray}
I_{i}^{net} &=& \sum\limits_{j \in J_{exc}(i)} g^{net} r_j(v_i - E_{ex}) + \sum\limits_{j \in J_{inh}(i)} g^{inh} r_j(v_i - E_{in}) \label{eq:Innet}
\end{eqnarray}
Here, $J_{exc}(i)$ is the set of cells directly projecting excitatorily to cell $i$ ($J_{inh}(i)$: inhibitorily). The parameter $g^{net}$ is the maximum conductance of network-internal (lateral) synapses, marked as excitatory by the synaptic reversal potential $E_{ex} = 0$. Similarly, $g^{inh}$ scales the conductance of inhibitory synapses, with reversal potential $E_{in} = -80$. Each synaptic conductance is further modulated by the fraction of open receptors $r_j$, which varies in accordance with incoming spikes. Specifically, $r_j$ is driven by the concentration of neurotransmitter in the synaptic cleft $[T]_j$, which, in turn, is a pulse of duration $\tau=0.02$ after each incoming spike:
\begin{eqnarray}
\dot{r_j} &=& \alpha [T]_j (1-r_j) - \beta r_j \label{eq:dr} \\
\left[T\right]_j &=& \Theta (T_j + \tau - t)\, \Theta (t-T_j) \label{eq:Tj}
\end{eqnarray}
$\dot{r_i}$ is parameterized by the rise and decay time constants $\alpha = \beta = 8$. The transmitter concentration $[T]_j$ is given by the product of two heaviside step functions $\Theta$, chosen such that transmitter is present ($[T]_j=1$) precisely from time $T_j$ at which a presynaptic spike occured until time $T_j + \tau$. Upstream connections -- responsible for delivering external, random input to the network -- can be either excitatory or inhibitory (parameterization below). This results in a slightly more complex, but essentially similar formulation for the input current $I_i^{up}$. To allow for excitatory and inhibitory input synapses, we introduce a number of additional terms: The conductances of excitatory and inhibitory upstream synapses $g^{up,ex}$ and $g^{up,in}$, corresponding terms for the fractions of open receptors $r_i^{up,ex}$ and $r_i^{up,in}$, for transmitter concentrations $[T]_i^{up,ex}$ and $[T]_i^{up,in}$, and finally for the arrival times of spikes on excitatory and inhibitory input synapses, $T_i^{up,ex}$ and $T_i^{up,in}$.
\begin{eqnarray}
I_i^{up} &=& g^{up,ex} r_i^{up,ex}(v_i - E_{ex}) \;+\; g^{up,in} r_i^{up,in}(v_i - E_{in}) \label{eq:Iup} \\
\dot{r_i}^{up,ex} &=& \alpha [T]_i^{up,ex} (1-r_i^{up,ex}) - \beta r_i^{up,ex} \label{eq:drupex} \\
\dot{r_i}^{up,in} &=& \alpha [T]_i^{up,in} (1-r_i^{up,in}) - \beta r_i^{up,in} \label{eq:drupin} \\
\left[T\right]_i^{up,ex} &=& T_{max}\; \Theta (T_i^{up,ex} + \tau - t)\, \Theta (t-T_i^{up,ex}) \label{eq:Tiupex} \\
\left[T\right]_i^{up,in} &=& T_{max}\; \Theta (T_i^{up,in} + \tau - t)\, \Theta (t-T_i^{up,in}) \label{eq:Tiupin}
\end{eqnarray}
The random excitatory and inhibitory input spikes occur independently of each other and across time and space. Concretely, if neuron $i$ is set to receive external input, the number of spike events per ms is Poisson distributed with rates $\lambda_{ex} = 40$ and $\lambda_{in}=1$, respectively. This is realized by independent coin flips with success probabilities $\lambda_{ex} \cdot \Delta$ and $\lambda_{in} \cdot \Delta$ at each numerical integration step, of which $\Delta^{-1} = 200$ are performed per ms.

\subsection*{Measuring synchrony}

Throughout the article, we measure the average degree of zero-lag synchrony of a particular subpopulation in the network, following \citet{jordistrogatz}. Spiketrains from each neuron are convolved with a causal exponential kernel $k(t) = e^{2t}$, yielding an activation trace $A_i$ per neuron. With this, the synchrony of a population $S$ during an interval $T$ is given by the variance of the mean field of $S$, normalized by the avarage variance of the members of $S$.

    $$ R_{syn}(S, T) = \frac{\widehat{\mathrm{Var}}[\langle A_i(t)\rangle_{i \in S}]_{t \in T}}  {\langle \widehat{\mathrm{Var}}[A_i(t)]_{t \in T} \rangle_{i \in S}}$$

Intuitively, if all members of the measured population fire strictly simultaneously, the mean activity of these cells fluctuates just as strongly as each of their individual activities (producing a value of $R_{syn} = 1$). On the other hand, if cells fire out of phase, their mean activity is comparably stable, while each individual cell still fluctuates as much as before, leading to a lower, though nonzero value.

\subsection*{Measuring synchrony differences over time}

To quantify for how long a network needs to be observed until a difference in synchrony between stimuli becomes apparent, we take synchrony measurements in increasingly long time intervals. The earlier a correct classification by stimulus is possible, the more quickly the degrees of synchrony of these stimuli must have diverged.
In more technical terms, we perform repeated, independent Gaussian naive Bayes classifications on synchrony values measured in growing window increments, as follows.\\
For each stimulus condition, a copy of the network is driven by a given input pattern and voltage traces of a subpopulation $S$ in this copy are recorded. Each stimulus condition is thus identified with an independent, separately measured population $S$. In each population or condition $S$ and each window increment step $t$, the synchrony value $x_{S,t} := R_{syn}(S,[0,t])$ is assumed to follow a normal distribution. Parameters $\mu_{S,t}$ and $\sigma_{S,t}$ of each such distribution are estimated by the sample mean $\hat{\mu}_{S,t}$ and sample variance $\hat{\sigma}_{S,t}$ of synchrony values measured in half the available simulation trials. We arrive at an estimated density $l_{S,t}$ over possible synchrony values $x$ for each measured condition and at each window increment:

$$l_{S,t}(x) := \mathcal{N}(x|\hat{\mu}_{S,t} , \hat{\sigma}_{S,t})$$

Treating this as a likelihood function and assuming (discrete) uniform prior probabilities for the different conditions, the posterior probability that condition $S$ is the origin of some newly measured synchrony value $x$ at window increment $t$ is therefore given by

$$P(S|x,t) = \frac{l_{S,t}(x) P(S)}  {\sum\limits_{S'}l_{S',t}(x) P(S')} $$

with $S'$ iterating all considered stimulus conditions. Finding out how long a given trial needs to be observed until it can be classified, thus, amounts to finding the window length $t$ after which this series of posterior probabilities crosses the decision threshold ($0.5$ in our case of two stimuli). This duration -- or rather, the distribution of such durations across many simulation trials -- is reported. More precisely, we report the distribution of number of spikes fired up to that time, on average over the measured set of cells.

\subsection*{Probabilistic formulation of familiarity}
Above, we used an intuitive description of pattern familiarity, namely the fraction of cells in a given pattern that overlap with the network's input history, which is a set $H$ of patterns. This can equivalently be expressed in probabilistic terms. Let the prior probability of occurence of an input pattern (set of active cells) $S=\{c_0,...c_n\}$ be given by a joint event of individual activations of its members:
$$P(S) = P(\bigcup_{c\in S}c) = \sum_{c\in S} P(c)$$
The main simplifying assumption of the discussed familiarity measure is that these individual activation probabilities $P(c)$ are binarized, in the sense that a constant, non-zero activation probability is assigned only to cells that appear in patterns found in $H$:
$$P(c) = \mathbbm{1}_{\bigcup H}(c)|S|^{-1}$$
With the choice of normalization $|S|$, patterns $S$ that fall fully within $H$ have prior probability $P(S)=1$, whereas those which miss $H$ altogether have probability 0, with intermediate values for partial overlaps. It is in this precise sense that we call synchrony an estimate of the prior probability of an incoming stimulus pattern.\\

\subsection*{Code Availability}
Annotated source files are found at {\url{https://github.com/cknd/synchrony}}.

\section*{Acknowledgments}
J.G.O. was supported by the Ministerio de Economia y Competividad and FEDER
(Spain, project FIS2015-66503-C3-1-P) and the ICREA Academia programme.\\

E.U. acknowledges support from the Scottish Universities Life Sciences Alliance (SULSA) and HPC-Europa2.\\

\end{document}